\documentclass{ws-procs9x6}

\begin{document}

\title{Particles and propagators in Lorentz-violating supergravity}
\author{Roland E. Allen and Seiichirou Yokoo}
\maketitle

\address{Physics Department, Texas A\&M University \\
College Station, TX 77843, USA\\
email: allen@tamu.edu}

\abstracts{We obtain the propagators for spin 1/2 fermions and sfermions in
Lorentz-violating supergravity. }

Any violation of Lorentz invariance must be extremely small for ordinary
matter under ordinary conditions\cite{kostelecky}. However, in
Lorentz-violating supergravity\cite{allen-2003} there is Lorentz violation
for both Standard Model particles at very high energy\cite{allen-yokoo-2003}
and their supersymmetric partners at even relatively low energy\cite
{allen-2002}. Here w{e obtain the propagators for the fermions and 
sfermions }of this theory, with the prefix ``s'' standing for
``supersymmetric partner'' rather than ``scalar'' in the present context,
since these particles are spin 1/2 rather than spin zero bosons\cite
{allen-2002}.

In the present paper we also extend our previous work by considering 
left-handed as well as right-handed sfermion fields. 
Initially all fermion fields are right-handed, but
one can transform half of them to left-handed fields\cite{allen-yokoo-2003},
obtaining, e.g., the full 4-component field $\psi $ for the electron. The
same transformation can be employed for the spin $1/2$ sfermions of the
present theory, with only one change: In the fourth step leading up to (11)
in Ref. 3, bosonic fields commute rather than anticommute, so the final
Lagrangian for massless particles has the form
\begin{equation}
\mathcal{L}_{L}=\pm \frac{1}{2}\,\left( \bar{m}^{-1}\eta ^{\mu \nu }\partial 
_{\mu }\psi _{L}^{\dagger }\partial _{\nu }\psi _{L}+\psi _{L}^{\dagger }i
\bar{\sigma}^{\mu }\partial _{\mu }\psi _{L}\right) +h.c.
\end{equation}
where the upper sign holds for fermions and the lower for bosons. Here $\psi
_{L}$ is a 2-component left-handed spinor, with 
$\bar{\sigma}^{k} = - \sigma^{k} $ as usual, and $\eta ^{\mu \nu
}=diag(-1,1,1,1)$. The total Lagrangian has the following form, 
with left- and right-handed fields combined in a 4-component spinor $\psi $ 
(in the Weyl representation, and coupled by a Dirac mass $m$ in the case 
of fermions):
\begin{equation}
\mathcal{L}_{\psi }=\bar{m}^{-1}\dot{\psi}^{\dagger }\widetilde{\gamma }_{2}
\dot{\psi}+\left( \frac{1}{2}i\psi ^{\dagger }\widetilde{\gamma }_{1}
\dot{\psi}+h.c.\right) +\mathcal{L}_{\psi }^{\prime }
\end{equation}
where $\widetilde{\gamma }_{1}$ and $\widetilde{\gamma }_{2}$ are 
diagonal $4\times 4$ matrices (with elements $\pm 1$) inserted to cover 
all the sign possibilities.

We treat both Standard Model fermions and bosonic sfermions together, with
the following conventions: (1) The canonical momenta conjugate to $\psi $
and $\psi ^{\dagger }$ are respectively called $\pi ^{\dagger }$ and 
$\tilde{\pi} $.
(2) In defining these momenta, the derivative is taken from the right. The 
momenta are then 
\begin{equation}
\pi ^{\dagger }=\frac{\partial \mathcal{L}_{\psi }}{\partial \dot{\psi}}=\;
\bar{m}^{-1}\dot{\psi}^{\dagger }\widetilde{\gamma }_{2}+\frac{1}{2}i\psi
^{\dagger }\widetilde{\gamma }_{1}\quad ,\quad 
\tilde{\pi} =\frac{\partial \mathcal{L}
_{\psi }}{\partial \dot{\psi}^{\dagger }}= \mp \bar{m}^{-1}\widetilde{\gamma }
_{2}\dot{\psi} \pm \frac{1}{2}i\widetilde{\gamma }_{1}\psi .
\end{equation}
The equation of motion has the form 
\begin{equation}
-\widetilde{\gamma }_{2}\bar{m}^{-1}\frac{\partial ^{2}}{\partial t^{2}}\psi
+i\widetilde{\gamma }_{1}\frac{\partial }{\partial t}\psi -H^{\prime }\psi =0
\end{equation}
and we quantize by requiring that 
\begin{align}
\left[ \psi _{\alpha }\left( \vec{x},x^{0}\right) ,\pi _{\beta }^{\dagger
}\left( \vec{x}\,^{\prime },x^{0}\right) \right] _{\pm }
& =i\delta \left( \vec{x}-\vec{x}\,^{\prime }\right) \delta _{\alpha \beta } \\ 
\left[ \psi _{\alpha }^{\dagger }\left( \vec{x},x^{0} \right) , 
\tilde{\pi} _{\beta }\left( \vec{x}\,^{\prime },x^{0} \right) \right] _{\pm } 
& =i\delta \left( \vec{x}-\vec{x}\,^{\prime }\right) \delta _{\alpha \beta } .
\end{align}
The retarded Green's function is defined by 
\begin{equation}
iG_{\alpha \beta }^{R}\left( x,x^{\prime }\right) =\theta \left( t-t^{\prime
}\right) \left\langle 0\left| \left[ \psi _{\alpha }\left( x\right) ,\psi
_{\beta }^{\dagger }\left( x^{\prime }\right) \right] _{\pm }\right|
0\right\rangle 
\end{equation}
and we can show that it is in fact a Green's function by using 
\begin{align}
\,\frac{\partial ^{2}}{\partial t^{2}}\left( \theta \left( t-t^{\prime
}\right) f\left( t\right) \right) & =\frac{\partial \delta \left(
t-t^{\prime }\right) }{\partial t}f\left( t\right) +2\delta \left(
t-t^{\prime }\right) \frac{f\left( t\right) }{\partial t}+\theta \left(
t-t^{\prime }\right) \frac{\partial ^{2}f\left( t\right) }{\partial t^{2}} 
\nonumber \\
& =\delta \left( t-t^{\prime }\right) \frac{\partial f\left( t\right) }
{\partial t}+\theta \left( t-t^{\prime }\right) \frac{\partial ^{2}f\left(
t\right) }{\partial t^{2}}
\end{align}
to obtain 
\begin{align}
& i\left( -\widetilde{\gamma }_{2}\bar{m}^{-1}\frac{\partial ^{2}}{\partial 
t^{2}}+i\widetilde{\gamma }_{1}\frac{\partial }{\partial t}-H^{\prime
}\right) G_{\alpha \beta }^{R}\left( x,x^{\prime }\right)   \nonumber \\
& \hspace{0.5cm}=\delta \left( t-t^{\prime }\right) \left( -\widetilde{
\gamma }_{2}\bar{m}^{-1}\frac{\partial }{\partial t}+i\widetilde{\gamma }
_{1}\right) \left\langle 0 \left| \left[ \psi _{\alpha }\left( x\right) ,\psi
_{\beta }^{\dagger }\left( x^{\prime }\right) \right] _{\pm } 
\right| 0\right\rangle \nonumber \\
& \hspace{1cm}+\theta \left( t-t^{\prime }\right) \left( -\widetilde{\gamma }
_{2}\bar{m}^{-1}\frac{\partial ^{2}}{\partial t^{2}}+i\widetilde{\gamma }_{1}
\frac{\partial }{\partial t}-H^{\prime }\right) \left\langle 0 \left| \left[ 
\psi_{\alpha }\left( x\right) ,\psi _{\beta }^{\dagger }\left( x^{\prime
}\right) \right] _{\pm } \right| 0\right\rangle   \nonumber \\
& \hspace{0.5cm}=2\delta \left( t-t^{\prime }\right) \left\langle 0 
\left| \left[ \psi _{\beta }^{\dagger }\left( x^{\prime}\right) , 
\tilde{\pi} _{\alpha }\left( x\right) \right] _{\pm } 
\right| 0\right\rangle +O\left( \bar{m}^{-1} \right) +0  \nonumber \\
& \hspace{0.5cm}=2i\delta ^{\left( 4\right) }\left( x-x^{\prime }\right)
\delta _{\alpha \beta }+O\left( \bar{m}^{-1} \right) 
\end{align}
where $O\left( \bar{m}^{-1} \right) $ represents a term which will 
become negligibly small at energies low compared to $\bar{m}$, after 
extremely high energy terms have been discarded from the representation 
of $G^{R}$. (See the discussion below (30).) The causal Green's function 
is defined by 
\begin{align}
iG_{\alpha \beta }\left( x,x^{\prime }\right) & =\left\langle 0\left| T\left(
\psi _{\alpha }\left( x\right) \psi _{\beta }^{\dagger }\left( x^{\prime
}\right) \right) \right| 0\right\rangle \\
& =iG_{\alpha \beta }^{R}\left(
x,x^{\prime }\right) \mp \left\langle 0 \left| \psi _{\beta }^{\dagger }\left(
x^{\prime }\right) \psi _{\alpha }\left( x\right) \right| 0\right\rangle 
\end{align}
so it satisfies the equation 
\begin{align}
& \left( -\widetilde{\gamma }_{2}\bar{m}^{-1}\frac{\partial ^{2}}{\partial 
t^{2}}+i\widetilde{\gamma }_{1}\frac{\partial }{\partial t}-H^{\prime
}\right)  G\left (x,x^{\prime }\right) \nonumber \\
& = \left( -\widetilde{\gamma }_{2}\bar{m}^{-1}\frac{\partial ^{2}}{\partial 
t^{2}}+i\widetilde{\gamma }_{1}\frac{\partial }{\partial t}-H^{\prime
}\right) G^{R}\left( x,x^{\prime }\right) +0 \nonumber \\
& = 2\delta ^{\left( 4\right) }\left( x - x^{\prime }\right) 
+O\left( \bar{m}^{-1} \right) 
\end{align}
where a $4\times 4$ identity matrix implicitly multiplies 
$\delta ^{\left( 4\right) }\left( x - x^{\prime }\right)$
and the factor of $2$ will be explained below.

Let $\psi _{n}$ and $\psi _{m}$ respectively represent the
positive-frequency and negative-frequency solutions to (4). With 
$b_{m}^{\dagger }=a_{m}$, the field can be represented as in (4.17), (4.45),
and (4.55) of Ref. 4: 
\begin{equation}
\psi =\sum_{n}\,a_{n}\psi _{n}+\sum_{m}\,b_{m}^{\dagger }\psi _{m}
\end{equation}
with
\begin{align}
\psi _{n}\left( x\right) & =A_{\lambda }\left( p^{\prime }\right) u_{\lambda
}\left( p^{\prime }\right) \exp \left( -i\varepsilon _{\lambda }\left( 
\vec{p}\,^{\prime }\right) t\right) \exp \left( i\vec{p}\,^{\prime }\cdot 
\vec{x}\right) \\
\psi _{m}\left( x\right) & =A_{\kappa }\left( p^{\prime }\right) v_{\kappa
}\left( p^{\prime }\right) \exp \left( +i\varepsilon _{\kappa }\left( \vec{p}
\,^{\prime }\right) t\right) \exp \left( -i\vec{p}\,^{\prime }\cdot \vec{x}
\right)
\end{align}
so that $n\leftrightarrow \vec{p}\,^{\prime },\lambda $ and 
$m\leftrightarrow -\vec{p}\,^{\prime },\kappa $. Here $u$ and $v$ are
4-component spinors, and the normalization is the same as in (4.35) of Ref.
4: 
\begin{eqnarray}
A_{\lambda }^{*}\left( p^{\prime }\right) A_{\lambda }\left( p^{\prime
}\right) &=&\left( 1+2\varepsilon _{\lambda }\left( \vec{p}\,^{\prime
}\right) /\bar{m}\right) ^{-1}V^{-1} \\
A_{\kappa }^{*}\left( p^{\prime }\right) A_{\kappa }\left( p^{\prime
}\right) &=&\left( 1-2\varepsilon _{\kappa }\left( \vec{p}\,^{\prime
}\right) /\bar{m}\right) ^{-1}V^{-1}.
\end{eqnarray}

To obtain the Green's function we need 
\begin{align}
\left\langle 0|\psi \left( x\right) \psi ^{\dagger }\left( x^{\prime
}\right) |0\right\rangle & =\left\langle 0\left| \sum_{nn^{\prime }}\psi
_{n}\left( x\right) \psi _{n^{\prime }}^{\dagger }\left( x^{\prime }\right)
\,\left( \delta _{nn^{\prime }}\mp a_{n^{\prime }}^{\dagger }a_{n}\,\right)
\right| 0\right\rangle \\
& =\sum_{n}\psi _{n}\left( x\right) \psi _{n}^{\dagger }\left( x^{\prime
}\right) \\
\left\langle 0|\psi ^{\dagger }\left( x^{\prime }\right) \psi \left(
x\right) |0\right\rangle & =\sum_{m}\psi _{m}\left( x\right) \psi
_{m}^{\dagger }\left( x^{\prime }\right) .
\end{align}
The Fourier transform of the causal Green's function 
\begin{equation}
G\left( \omega, \vec{p} \right) =\int dt\,\,
\exp \left( i\omega \left( t-t^{\prime }\right)
\right) \int d^{3}x\,\,\exp \left( -i\vec{p}\cdot \left( \vec{x}-\vec{x}
\,^{\prime }\right) \right) G\left( x,x^{\prime }\right)   \nonumber
\end{equation}
can be found by using 
\begin{align}
& \theta \left( t-t^{\prime }\right)  =\int \frac{d\omega ^{\prime }}{2\pi i}
\,\frac{\exp \left( i\omega ^{\prime }\left( t-t^{\prime }\right) \right) }
{\omega ^{\prime }-i\epsilon } \\
& \int d^{4}x\,\exp \left( i\left( \vec{p}\,^{\prime }-\vec{p}\right) \cdot 
\left( \vec{x}-\vec{x}\,^{\prime }\right) \right)  = V\,\delta _{\vec{p}
\vec{p}\,^{\prime }}
\end{align}
in each of the two terms: 
\begin{align}
& \int dt\,\exp \left( i\omega \left( t-t^{\prime }\right) \right) \,\int
d^{3}x\,\,\exp \left( -i\vec{p}\cdot \left( \vec{x}-\vec{x}\,^{\prime
}\right) \right) \theta \left( t-t^{\prime }\right) \sum_{n}\psi
_{n}\left( x\right) \psi _{n}^{\dagger }\left( x^{\prime }\right) \nonumber \\
& =\,\int \frac{d\omega ^{\prime }}{2\pi i}\frac{2\pi }{\omega ^{\prime
}-i\epsilon }\,\delta \,\left( \omega +\omega ^{\prime }-\varepsilon
_{\lambda }\left( \vec{p}\right) \right) \sum\limits_{\vec{p}^{\prime
}\lambda }\,A_{\lambda }^{*}\left( p^{\prime }\right) A_{\lambda }\left(
p^{\prime }\right) u_{\lambda }\left( p^{\prime }\right) u_{\lambda
}^{\dagger }\,\left( p^{\prime }\right) V\delta _{\vec{p}\vec{p}\,^{\prime
}}\,  \nonumber \\
& =i\sum\limits_{\lambda }\,\,\frac{u_{\lambda }\left( {\vec{p}}
,p^{0}\right) u_{\lambda }^{\dagger }\left( \vec{p},p^{0}\right) }{\omega
-\varepsilon _{\lambda }\left( \vec{p}\right) +i\epsilon }\,\frac{1}{\left(
1+2\varepsilon _{\lambda }\left( \vec{p} \right) /\bar{m}\right) }
\quad , \quad p^{0}=\varepsilon \left( \vec{p}\right) 
\end{align}
and 
\begin{align}
& \int dt\,\exp \left( i\omega \left( t-t^{\prime }\right) \right) \int
d^{3}x\,\,\exp \left( -i\vec{p}\cdot \left( \vec{x}-\vec{x}\,^{\prime
}\right) \right) \theta \left( t^{\prime }-t\right) \sum_{m}\psi
_{m}\left( x\right) \psi _{m}^{\dagger }\left( x^{\prime }\right) \nonumber \\
& =\,\int \frac{d\omega ^{\prime }}{2\pi i}\frac{2\pi \,}{\omega ^{\prime
}-i\epsilon }\delta \,\left( \omega -\omega ^{\prime }+\varepsilon _{\kappa
}\left( \vec{p}\right) \right) \sum\limits_{\vec{p}^{\prime }\kappa
}\,A_{\kappa }^{*}\left( p^{\prime }\right) A_{\kappa }\left( p^{\prime
}\right) v_{\kappa }\left( p^{\prime }\right) v_{\kappa }^{\dagger }\,\left(
p^{\prime }\right) \,V\delta _{\vec{p},-\vec{p}\,^{\prime }}  \nonumber \\
& =-i\sum\limits_{\kappa }\,\,\frac{v_{\kappa }\left( -\vec{p},-p^{0}\right)
v_{\kappa }^{\dagger }\,\left( -\vec{p},-p^{0}\right) }{\omega +\varepsilon
_{\kappa }\left( \vec{p}\right) -i\epsilon }\,\frac{1}{\left( 1-2\varepsilon
_{\kappa }\left( \vec{p} \right) /\bar{m}\right) }
\quad , \quad -p^{0}=\varepsilon \left( \vec{p}\right) 
\end{align}
since $\varepsilon \left( -\vec{p}\right)=\varepsilon \left(\vec{p}\right)$. 
When combined these expressions give 
\begin{eqnarray}
G\left( \omega, \vec{p} \right)  &=&\frac{1}{2}\sum\limits_{\lambda }\,\,
\frac{u_{\lambda
}\left( p\right) u_{\lambda }^{\dagger }\left( p\right) }{\omega
-\varepsilon _{\lambda }\left( \vec{p}\right) +i\epsilon }\;\frac{1}{
1+2\varepsilon _{\lambda }\left( \vec{p} \right) /\bar{m}} \nonumber \\
&&\mp \frac{1}{2}\sum\limits_{\kappa }\,\,\frac{v_{\kappa }\left( -p\right)
v_{\kappa }^{\dagger }\,\left( -p\right) }{\omega +\varepsilon _{\kappa
}\left( \vec{p}\right) -i\epsilon }\;\frac{1}{1-2\varepsilon _{\kappa
}\left( \vec{p} \right) /\bar{m}}
\end{eqnarray}
with $u_{\lambda }\left( p\right) u_{\lambda }^{\dagger }\left( p\right)$ 
and $v_{\kappa }\left( -p\right) v_{\kappa }^{\dagger }\,\left(-p\right)$ 
on the energy shell, in the sense that $p^{0}=\varepsilon
\left( \vec{p}\right) $ in the first term and $-p^{0}=\varepsilon \left( 
\vec{p}\right) $ in the second. However, $G\left( x-x^{\prime }\right) $ can
be equally well represented by 
\begin{equation}
G\left( x-x^{\prime }\right) =\int \frac{d p^{0} }{2\pi }\,\exp \left(
-i p^{0} \left( t-t^{\prime }\right) \right) \;\sum\limits_{\vec{p}}\exp
\left( i\vec{p}\cdot \left( \vec{x}-\vec{x}\,^{\prime }\right) \right) 
\tilde{G}\left( p\right)  
\end{equation}
\begin{align}
& \tilde{G}\left( p\right)  \nonumber \\
& = \frac{1}{2}\sum\limits_{\lambda }\,\,
\frac{u_{\lambda }\left( p\right) u_{\lambda }^{\dagger }\left( p\right) }
{p^{0}-\varepsilon _{\lambda }\left( \vec{p}\right) +i\epsilon }\;\frac{1}
{1+2p^{0}/\bar{m}} \mp \frac{1}{2}\sum\limits_{\kappa }\,\,\frac{v_{\kappa }
\left( -p\right)
v_{\kappa }^{\dagger }\,\left( -p\right) }{p^{0}+\varepsilon _{\kappa
}\left( \vec{p}\right) -i\epsilon }\;\frac{1}{1-2p^{0}/\bar{m}} 
\end{align}
with $p^{0}$ unrestricted, since, when the residue is evaluated at one of the 
poles, $p^{0}$ is forced to equal 
$\pm \left ( \varepsilon \left ( \vec{p} \right ) - i \epsilon \right )$.

Let us define a modified Green's function $\tilde{S}_{f}$ for a fermion by 
\begin{equation}
\tilde{S}_{f}\left( x,x^{\prime }\right) =\left\langle 0\left| T\left( \psi
\left( x\right) \bar{\psi}\left( x^{\prime }\right) \right) \right|
0\right\rangle \quad ,\quad \bar{\psi}=\psi ^{\dagger }\gamma ^{0} 
\end{equation}
\begin{align}
& \tilde{S}_{f}\left (p \right) =i\tilde{G}\left( p\right) \gamma ^{0} \\
& =\frac{i}{2}\sum\limits_{\lambda }\,\,\frac{u_{\lambda }\left( p\right) 
\bar{u}_{\lambda }\left( p\right) }{p^{0}-\varepsilon _{\lambda }\left( 
\vec{p}\right) +i\epsilon }\;\frac{1}{1+2p^{0}/\bar{m}} 
-\frac{i}{2}\sum\limits_{\kappa }\,\,\frac{v_{\kappa }\left( -p\right) 
\bar{v}_{\kappa }\left( -p\right) }{p^{0}+\varepsilon _{\kappa }\left( 
\vec{p}\right) -i\epsilon }\;\frac{1}{1-2p^{0}/\bar{m}}.
\end{align}
In the remainder of this paper we limit attention to energies that are low
compared to $\bar{m}$. In this case, and for massless 
right-handed fermions, it can be 
seen in (4.18)-(4.21) of Ref. 4 that there is only one value of $\lambda $,
and it corresponds to the normal branch with $\varepsilon _{\lambda }\left( 
\vec{p}\right) =\left| \vec{p}\right| $. On the other hand, there are three
values of $\kappa $: One corresponds to the normal branch with $\varepsilon
_{\kappa }\left( \vec{p}\right) =\left| \vec{p}\right| $, and two to
extremely high energy branches with $\varepsilon _{\kappa }\left( \vec{p}
\right) =\bar{m}\pm \left| \vec{p}\right| $. When a Dirac mass is
introduced, these last two branches are hardly perturbed, and they will
still give extremely large denominators in the expression above for 
$\tilde{S}_{f}$. They can then be neglected in calculations at normal 
energies, and at the same time $2p^{0}/\bar{m}$ can be neglected. With 
the high energy branches omitted, we have the relevant low energy propagator 
\begin{equation}
S_{f}\left( p\right) 
=\frac{i}{2}\sum\limits_{\lambda=1,2 }\,\frac{u_{\lambda }\left(
p\right) \bar{u}_{\lambda }\left( p\right) }{p^{0}-\varepsilon _{\lambda
}\left( \vec{p}\right) +i\epsilon }-\frac{i}{2}\sum\limits_{\kappa=1,2}
\,\frac{v_{\kappa }\left( -p\right) \bar{v}_{\kappa }\left(
-p\right) }{p^{0}+\varepsilon _{\kappa }\left( \vec{p}\right) -i\epsilon }
\end{equation}
where $\varepsilon \left( \vec{p}\right) =\left( \vec{p}\,^{2}+m^{2}
\right) ^{1/2}$. 
In the sums, $\lambda $ and $\kappa $ are now 
each limited to the $2$ usual values for a $4$-component Dirac spinor, 
rather than the total of $8$ values that one would have if the $4$
extremely high energy solutions were retained. We should note, however,
that the high-energy solutions give a contribution in the equation of motion
(9) for the Green's function that is equal to that of the low-energy
solutions, because the derivatives bring down large energies which cancel
those in the denominator. This accounts for the factor of $2$ in (9) 
and (12), and at normal energies we obtain 
\begin{eqnarray}
\left( \rlap/p-m\right) iS_{f}\left( x,x^{\prime
}\right)  = \delta ^{\left( 4\right) }\left( x - x^{\prime
}\right)
\end{eqnarray}
where $\rlap/p=-\gamma ^{\mu }\partial _{\mu }$ with our metric tensor $\eta
_{\mu \nu }=diag\left( -1,1,1,1\right) $. The usual Dirac spinors 
$u_{\lambda }^{D}$ and $v_{\kappa }^{D}$ have the completeness relation 
\begin{equation}
\sum\limits_{\lambda=1,2 }\,u_{\lambda }^{D}\left( p\right) 
\bar{u}_{\lambda }^{D}\left( p\right) \,=\rlap/p+m\quad ,\quad 
\sum\limits_{\kappa= 1,2 } \,v_{\kappa }^{D}\left( p\right) 
\bar{v}_{\kappa }^{D}\left( p\right) \,\,=\rlap/p-m
\end{equation}
and they are normalized to $2p^{0}$, whereas our $u_{\lambda }$ and 
$v_{\kappa }$ are normalized to unity. We then have 
\begin{equation}
S_{f}\left( p\right) =\frac{i}{2}\frac{1}{2p^{0}}\,\,\frac{\rlap/p+m}
{p^{0}-\varepsilon \left( \vec{p}\right) +i\epsilon }-\frac{i}{2}\frac{1}
{2p^{0}}\,\,\frac{-\rlap/p-m}{p^{0}+\varepsilon \left( \vec{p}\right)
-i\epsilon } 
=\,\frac{i}{\rlap/p-m+i\epsilon }
\end{equation}
and the standard expression for the Feynman propagator is regained.

For sfermions, however, one must make a distinction even at low energy
between the mathematical Green's function $\tilde{G}\left( p\right) $, in
which negative-norm solutions have been included, and the physical
propagator $S_{b}\left( p\right) $, which can contain only positive-norm
solutions. It is a fundamental requirement of quantum mechanics that
physical operators are allowed to connect only states in a positive-norm
Hilbert space. The physical field operator $\psi _{phys}$, and other
physical operators, should therefore contain only creation and destruction
operators for positive-norm states. To obtain the physical 
propagator, one can repeat the 
above treatment with only positive-norm solutions retained. In the present
theory, it fortunately turns out that one still has a complete set of
functions $\psi _{n}$ and $\psi _{m}$, as is required to provide a
proper representation of the original classical field and satisfy the
quantization condition (5).  On the other hand, one finds that 
the restriction to positive-norm solutions permits only one value each for 
$\lambda$ and $\kappa$ in (25) or (27) at low energy, corresponding to 
sfermions and anti-sfermions which are right-handed before a mass is 
introduced. The physical implications of this, and the issue of sfermion 
masses, will be discussed elsewhere.

\end{document}